\def\aa{{A\&A}}

\def\aj{{AJ}}
\def\annrev{{ARA\&A}}
\def\apj{{ApJ}}
\def\apjs{{ApJS}}

\def\mnras{{MNRAS}}
\def\nat{{Nature}}

\def\etal{et al.~}
\def\lax{{$\mathrel{\hbox{\rlap{\hbox{\lower4pt\hbox{$\sim$}}}\hbox{$<$}}}$}}
\def\gax{{$\mathrel{\hbox{\rlap{\hbox{\lower4pt\hbox{$\sim$}}}\hbox{$>$}}}$}}
\def\simpropto{{$\mathrel{\hbox{\rlap{\hbox{\lower4pt\hbox{$\sim$}}}\hbox{$\propto$}}}$}}

\documentclass[cup5b]{caps}
\usepackage{graphicx}
\usepackage{amssymb}
\usepackage{ociwsymp3}   

\HeadText{M. Donahue and G. M. Voit}

\begin{document}

\pagenumbering{arabic}

\author[]{MEGAN DONAHUE and G. MARK VOIT 
\\Space Telescope Science Institute, Baltimore, MD, USA
\\Michigan State University, East Lansing, MI, USA }

\chapter{Cool Gas in Clusters of Galaxies}

\begin{abstract}

Early X-ray observations suggested that the intracluster 
medium cools and condenses at the centers of clusters,
leading to a cooling flow of plasma in the cluster core.
The increased incidence of emission-line nebulosity,
excess blue light, AGN activity, and molecular gas
in the cores of clusters with short central cooling
times seemed to support this idea.  
However, high-resolution spectroscopic observations 
from {\em XMM-Newton} and {\em Chandra}
have conclusively ruled out simple, steady cooling flow
models.  We review the history of this subject, the current
status of X-ray observations, and some recent models
that have been proposed to explain why the core gas does not
simply cool and condense.

\end{abstract}

\section{A Census of Cool Gas}

Clusters of galaxies have very deep potential wells with
virial velocities equivalent to temperatures of $10^7 - 10^8$~K.
Gravitationally driven processes like accretion shocks and
adiabatic compression should therefore heat gas accumulating within 
a cluster to X-ray emitting temperatures.  Spectroscopic
X-ray observations show that most of a cluster's gas is indeed
near the virial temperature $T_{\rm vir} = \mu m_p \sigma^2_{\rm 1D} 
/ k$, equivalent to $7.1 \times 10^7 \, \sigma^2_{1000} \, {\rm K}$ or
$6.2 \, \sigma^2_{1000} \, {\rm keV}$, where $\sigma_{1000}$ is the
line-of-sight velocity dispersion in units of $1000 \, {\rm km \, s^{-1}}$
(Sarazin 1986).

Roughly 10\%--20\% of the baryons associated with clusters have
a temperature significantly less than the virial temperature,
qualifying as ``cool gas'' for the purposes of this review.
Much of this gas would be considered quite hot in other astrophysical
contexts, but in order to be cooler than the virial temperature
today, it must either have avoided the gravitational heating
experienced by the rest of the cluster or it must have significantly
cooled after entering the cluster.  

A large proportion of this cool gas is only moderately cooler 
than the virial temperature.  In the central $\sim$10\% of many 
clusters, corresponding to gas masses of $10^{11} - 10^{13} \, M_\odot$,
temperatures dip to $\sim T_{\rm vir}/2$.  Because this gas is dense
enough to radiate an energy equivalent to its thermal energy 
in less than a Hubble time, astronomers have long speculated 
that it cools and contracts, forming a ``cooling flow'' of
condensing gas in the cluster core (Cowie \& Binney 1977; Fabian \& Nulsen 
1977; Mathews \& Bregman 1978).

Gas much cooler than the virial temperature is also seen in clusters.
For example, all the stars in a cluster's galaxies are made of such
gas, implying that at least some cooling and condensation must
have occurred during the assembly of the cluster.  Applying
a standard mass-to-light ratio, one finds that $\sim 0.2 h^{3/2}$
of a cluster's baryons are ``cool gas'' of this kind (Arnaud et al. 1992;
White et al. 1993; with $h=H_0/100$ km s$^{-1}$ Mpc$^{-1}$).
While it may seem strange to include stars in a census of cool
intracluster gas, the total mass of stars does serve as a lower
limit to the amount of gas that passed through a cold phase
at some point in the cluster's past.

\begin{figure*}[t]
\includegraphics[width=1.00\columnwidth,angle=0,clip]{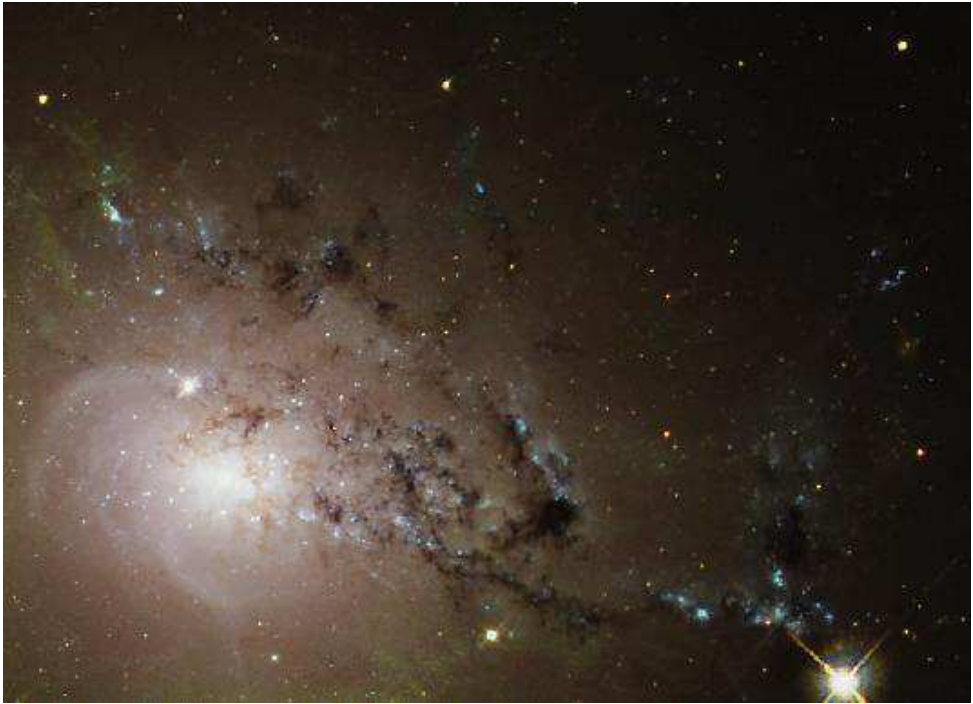}
\vskip 0pt \caption{
Hubble Heritage image of NGC~1275.
\label{fig-NGC1275}}
\end{figure*}

Many clusters also host optical emission-line nebulae within their
cores that appear to be associated with the cooler ($\sim T_{\rm vir}/2$)
X-ray emitting gas (Fabian \& Nulsen 1977; Ford \& Butcher 1979; Cowie et al. 
1983; Hu, Cowie, \& Wang 1985; Heckman et al. 1989; Crawford \& Fabian 1992;
Donahue, Stocke, \& Gioia 1992; Crawford 2003).  One could even say that 
Carnegie Observatories initiated the study of cool gas in clusters.  
Hubble \& Humason (1931) noted that NGC~1275, the central galaxy 
in the Perseus cluster, had a discrepant color index because of 
its strong emission spectrum, saying that ``it could be classified
as an elliptical nebula that has broken up without the formation
of spiral arms.''  Later, Baade \& Minkowski (1954)
noted that NGC~1275 was unusual among Seyfert galaxies because
its emission lines were not restricted to the nuclear regions.  
Lynds (1970) eventually imaged this amazing H$\alpha$ emission-line nebula 
using an interference filter.  Figure~\ref{fig-NGC1275} shows 
a recent Hubble Heritage close-up of NGC~1275, featuring a 
hint of spiral structure, complex dust lanes, and evidence 
for recent star formation.

The total amount of $\sim 10^4$~K gas in such nebulae is a mere 
$\sim 10^4 - 10^7 \, M_\odot$  (Heckman et al. 1989), but this nebulosity 
may be only the glowing skin surrounding considerably larger
masses of much cooler gas.  Clusters with H$\alpha$ emission also
have closely associated H$_2$ emission (Elston \& Maloney 1994;
Jaffe \& Bremer 1997; Falcke et al. 1998; Donahue et al. 2000; 
Jaffe, Bremer, \& van der Werf 2001; Edge et al. 2002).  Furthermore, recent
CO observations of a few cluster indicate that they may contain
up to $10^{9-11.5} \, M_\odot$ in the form of cool molecular gas
(Edge 2001).

The primary question concerning cool gas in clusters is whether
these pieces---cool X-ray gas, stars, nebulae, molecular 
clouds---all fit together into a single coherent picture of
condensation and star formation.  If so, then studies of cluster
cores may have much to teach us about the processes that govern
galaxy formation.  In this review, we will first recap the
cooling flow hypothesis, now over 25 years old, suggesting
that X-ray gas should cool and flow into cluster cores (see 
also Fabian, Nulsen, \& Canizares 1984, 1991; Fabian 1994).
Then we will present evidence showing that simple cooling
flows, in which cooling proceeds unopposed by heating or feedback,
do not occur (Molendi \& Pizzolato 2001; Peterson et al. 2001, 2003).
Supernovae and AGN activity must provide at
least some feedback during the history of the cluster.
In fact, {\em the global properties of clusters cannot
be understood without accounting for radiative cooling 
and subsequent feedback} (Lewis \etal 2000; 
Pearce et al. 2000; Voit \& Bryan 2001; Voit et al. 2002).  
Conduction may also suppress cooling in cluster cores (Bertschinger \&
Meiksin 1986; Bregman \& David 1988; Sparks, Macchetto, \& 
Golombek 1989), and this possibility has received renewed attention
in recent years (Malyshkin 2001; Narayan \& Medvedev 2001;  
Fabian, Voigt, \& Morris 2002; Voigt et al. 2002). 
However, we do not yet know which is the dominant mechanism 
opposing cooling---feedback, conduction, or perhaps a 
combination of the two (Ruskowkski \& Begelman 2002;
Brighenti \& Mathews 2003).
We close the review by summarizing a few clues that might
help answer this question.

\section{The Cooling Flow Hypothesis}

The road from the discovery of hot gas in clusters to
the cooling flow hypothesis was rather short.
Clusters of galaxies were first confirmed to be sources of 
X-ray emission in 1971 by the {\em UHURU} satellite (Gursky et al. 1971).
Thermal emission from hot intracluster gas seemed like a natural
interpretation (Lea et al. 1973; Lea 1975) given the extent of 
the emission (e.g., Forman et al. 1972; Kellogg et al. 1972) and the spectrum 
(e.g., Gorenstein et al. 1973; Davidsen et al. 1975; Kellogg, 
Baldwin, \& Koch 1975), but it was not confirmed until the
6.7~keV iron-line complex from helium-like and hydrogen-like ions
was discovered in the Perseus cluster by Mitchell et al. (1976)
using {\em Ariel V}, and in Virgo, Perseus, and Coma by 
Serlemitsos et al. (1977) using {\em OSO-8}.

Simple calculations of radiative cooling at the centers
of clusters like Perseus revealed that the cooling time, $t_c$,
was probably less than a Hubble time (Cowie \& Binney 1977; 
Fabian \& Nulsen 1977).  These authors suggested that, in the
absence of a compensating heat source, the core gas ought 
to cool and condense at the cluster's center.  Thus, the 
centers of all clusters with $t_c < H_0^{-1}$ soon became 
known as ``cooling flows,'' even though there was not yet
any firm evidence for either cooling or flowing.  The main
piece of circumstantial evidence was the close association
between a short central cooling time and the presence of
an optical emission-line nebula at the cluster's center,
presumed to be generated by gas cooling through $\sim 10^4$ K.
Hu et al. (1985) showed that these nebulae are
frequently found in clusters with $t_c \lesssim H_0^{-1}$, 
but never in clusters with $t_c > H_0^{-1}$.

A simple estimate of the implied cooling rate can be drawn
from the X-ray luminosity of the cooling region by assuming
the gas cools from the virial temperature at constant
pressure:
\begin{equation}
 \dot{M}_X \approx \frac {2} {5} \frac {\mu m_p} {k T_X} L_X(<r_c)
 \; \; .
\end{equation}
Here, $L_X(<r_c)$ is the X-ray luminosity coming from inside
the cooling radius $r_c$, at which $t_c \approx H_0^{-1}$.
Estimates for $\dot{M}$ derived from X-ray imaging often exceed
$100 \, M_\odot \, {\rm yr}^{-1}$ (Fabian et al. 1984), 
even approaching $1000 \, M_\odot \, {\rm yr}^{-1}$ 
in some extreme cases (e.g., White \etal 1994). 

The X-ray surface brightness distributions of cooling flow
clusters are inconsistent with steady flows in which 
$d\dot{M}/dr = 0$ because such flows produce exceedingly strong
central peaks in brightness.  To obtain better-fitting
surface brightness profiles, cooling flow modelers
allowed for spatially distributed mass deposition that
led to a decline in $\dot{M}$ as the flow approached
$r = 0$ (Fabian et al. 1981; Stewart et al. 1984). 
Models of this kind fit the data best if
$\dot{M}(r)$ \simpropto\ $r$ (Fabian et al. 1984),
implying that the flow must be inhomogeneous, 
with a range of cooling times at any given radius,
because only a subset of the inflowing gas manages
to condense within each radial interval 
(e.g., Thomas, Fabian, \& Nulsen 1987).
However, the overall $\dot{M}$ values derived from
such models are similar to the simple estimates
based on $L_X(<r_c)$.

Individual X-ray emission lines could, in principle, be used
to estimate the rate at which matter is cooling (Cowie 1981).
For cooling at constant pressure, the luminosity of emission line $i$
is
\begin{equation}
L_i = \dot{M} \frac{5k}{2\mu m}\int \frac{\epsilon_i(T)}{\Lambda(T)} dT,
\end{equation}
where $T$ is the plasma
temperature, $\epsilon_i(T)/\Lambda(T)$ is the fraction of the cooling
emissivity function owing to emission line $i$ as a function of $T$, 
$\mu m$ is the mean mass per
particle, and $k$ is the Boltzmann constant. In the steady
cooling flow model, this expression is integrated from $T=0$ to $T=T_{hi}$.
There were two high-resolution spectrometers on board the
{\em Einstein} Observatory, and results (with rather low
signal-to-noise ratio) from both of those spectrometers seemed to 
confirm the rates inferred from X-ray surface brightness distributions
(Canizares et al. 1982;
Canizares, Markert, \& Donahue 1988; Mushotzky \& Szymkowiak 1988).

\section{The Trouble with Cooling Flows}

X-ray astronomers have historically been quite fond of the
cooling flow hypothesis but have had trouble convincing
colleagues who work in other wavebands because no one
has ever found a central mass sink containing the
$\dot{M}_X H_0^{-1} \approx 10^{11-13}\, M_\odot$ implied by the
simplest interpretation of the X-ray observations.
Now that {\em Chandra} and {\em XMM-Newton} are providing
high-resolution spectra of cluster cores, X-ray astronomers
themselves have become convinced that cooling flows
are not that simple, if indeed they occur at all,
because the cooling rates derived from spectroscopy
do not agree with simple cooling flow predictions.

\subsection{The Mass-Sink Problem}

The trouble with cooling flows began when optical
observers could not locate all the stars that ought
to be formed in the prodigious cooling flows 
($> 100 \, M_\odot \, {\rm yr}^{-1}$) of some clusters (Fabian et al. 1991).
Star formation rates derived from observations of
excess blue light and H$\alpha$ nebulosity, assuming
a standard initial mass function, amounted to only 
$\lesssim 0.1 \dot{M}_X$ (Johnstone, Fabian, \& Nulsen 1987;
McNamara \& O'Connell 1992; Allen 1995; Cardiel, Gorgas, \& Arag\'on-Salamanca
et al. 1995, 1998).  While it remains
possible in principle that star formation in cooling
flows is heavily skewed toward unobservable low-mass
stars (Fabian, Nulsen, \& Canizares 1982), 
there is still no compelling theoretical 
justification for this idea.

Initial enthusiasm about the H$\alpha$ emission
representing $\sim 10^4$~K cooling flow gas (e.g., Cowie,
Fabian, \& Nulsen 1980) abated 
when it was realized that the $\dot{M}$ implied by 
the H$\alpha$ luminosity in some clusters was 
$\sim 10^2 \dot{M}_X$ (Cowie et al. 1983; Heckman
et al. 1989).  Models have been 
proposed in which the H$\alpha$ is boosted by
absorption of EUV and soft X-ray emission from cooling
gas (Voit \& Donahue 1990; Donahue \& Voit 1991)
or by cooling through turbulent mixing layers 
(Begelman \& Fabian 1990). 
However, it now seems likely that most of the
H$\alpha$ emission comes from photoionization
by OB stars (Johnstone et al. 1987; Voit \&
Donahue 1997; Cardiel et al. 1998; Crawford et al. 1999).

Hope for a solution to the mass-sink problem rose
with the apparent discovery of excess soft X-ray absorption
in cooling flow clusters, which would require $\sim 10^{12}\,
M_\odot$ of cold gas distributed over the central $\sim 100$~kpc
(White et al. 1991; Allen et al. 1993).  Yet, dogged pursuit of this 
cold gas by radio astronomers failed to find either 21~cm emission 
(Dwarakanath, van Gorkom, \& Owen 1994; O'Dea,
Gallimore, \& Baum 1995; O'Dea, Payne, \& Kocevski 1998)
or CO emission (O'Dea et al. 1994; Braine et al. 1995) with the 
necessary covering factor
and beam temperature.  Some clusters do have significant
amounts of molecular gas, but detections so far generally
find it only within the central $\sim 20$~kpc (Donahue
et al. 2000; Edge 2001; Edge et al. 2002).

One explanation for the undetectability of the cooling flow 
sink is that this gas may become so cold that it produces
no detectable emission (Ferland, Fabian, \& Johnstone 1994, 2002).
However, cold clouds bathed in the X-rays found in cluster
cores must reradiate the X-ray energy they absorb in
some other wave band.  At minimum, these clouds should
have an observable warm skin of detectable H~I if they
do indeed cover the central regions of clusters (Voit \& 
Donahue 1995).  Cold clouds with a low covering factor
may still evade current radio observations but would not 
produce appreciable soft X-ray absorption.

Soft X-ray absorption itself is probably now a phenomenon that
no longer needs explaining.  Recent cluster observations
with {\em Chandra} and {\em XMM-Newton} are failing
to confirm the levels of absorption suggested by
lower-resolution X-ray observations (McNamara et al. 2000;
Blanton, Sarazin, \& McNamara 2003; Peterson et al. 2003).  If these
observations are correct, then there is no evidence
at all, in any waveband, for a large mass sink in 
cooling flow clusters.

\begin{figure*}[t]
\includegraphics[width=1.00\columnwidth,angle=0,clip]{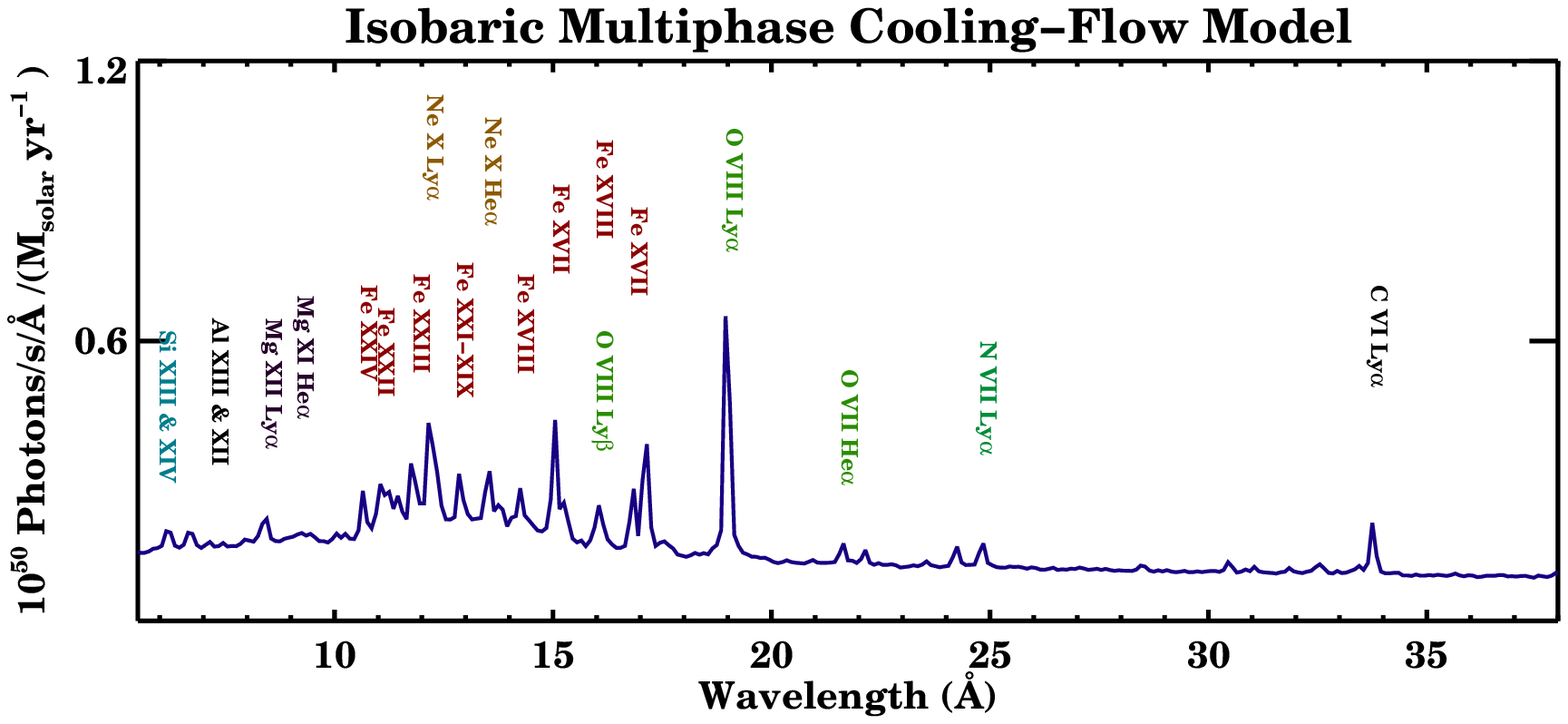}
\vskip 0pt \caption{
Spectrum emitted by gas cooling from 6~keV at
constant pressure.  Because the gas recombines
as it cools, the relative strengths of the
emission lines reveal how much gas cools
through each temperature. (Figure from Peterson et al. 2003.)
\label{fig-coolspec}}
\end{figure*}

\subsection{The Spectroscopic $\dot{M}$ Problem}

A recent breakthrough in X-ray astronomy is reframing
the whole debate about cooling flows.  In a simple,
steady-state cooling flow one expects to see emission
from gas over the entire range of temperature from
$T_{\rm vir}$ to the sink temperature, whatever
that may be.  Because the thermal energy lost as
gas cools from $T$ to $T-\Delta T$ is proportional
to $\Delta T$, the luminosity coming from gas within
that temperature interval is expected to be
$\Delta L \propto \dot{M} \Delta T$.  Thus, X-ray 
spectroscopy of the emission lines characteristic
of gas at each temperature can be used to test whether
$\Delta L / \Delta T$ is constant with temperature
(Cowie et al. 1980).  For example, we can use Fe XVII 
to track gas at $\lesssim 10^7$K, O VIII to track gas at 
$\lesssim 2 \times 10^7$, and UV observations of O~VI 
to track gas at $\sim 10^6$~K.  
Figure~\ref{fig-coolspec} shows the predicted spectrum if the cooling
gas is assumed to be an inhomogeneous (multiphase)
medium, as inferred from $\dot{M}$ \simpropto\ $r$, that 
cools at constant pressure.

\begin{figure*}[t]
\hspace{-1.5cm}
\includegraphics[width=1.10\columnwidth,angle=0,clip]{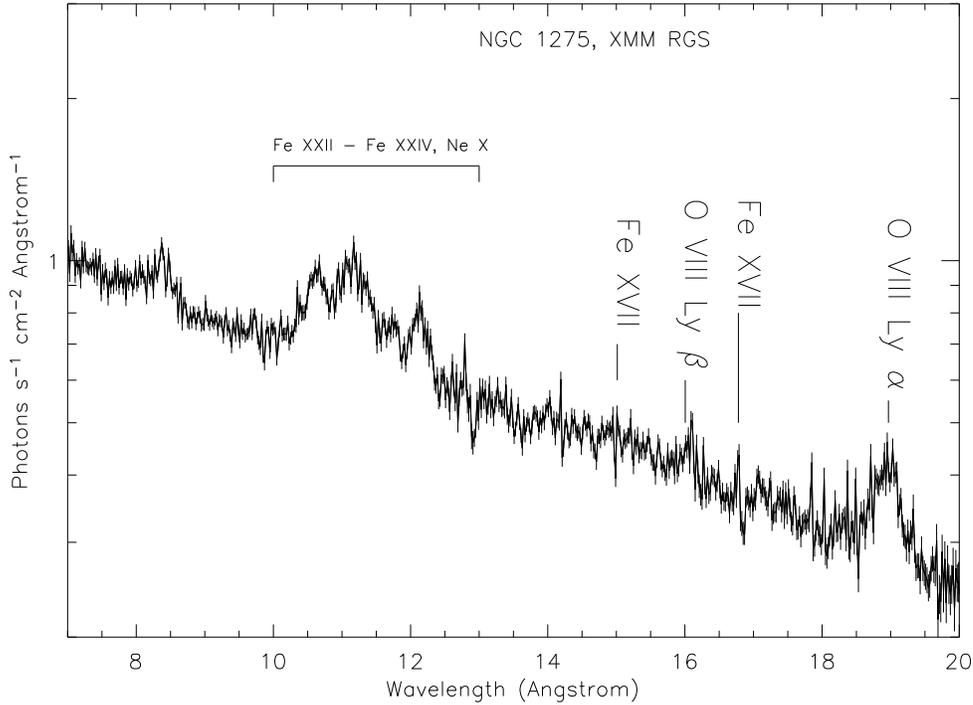}
\vskip 0pt \caption{
 Figure based on {\em XMM-Newton} RGS data for NGC 1275 in the Perseus
 cluster. The O~VIII Ly$\alpha$ and Ly$\beta$ lines were detected, but
 no Fe XVII is apparent at the expected wavelengths of 15.014 \AA\ or
 16.78 \AA.  (Data courtesy J. Peterson; Peterson et al. 2003.)
\label{fig-xmm-perseus}}
\end{figure*}

\begin{figure*}[t]
\hspace{-1.5cm}
\includegraphics[width=1.10\columnwidth,angle=0,clip]{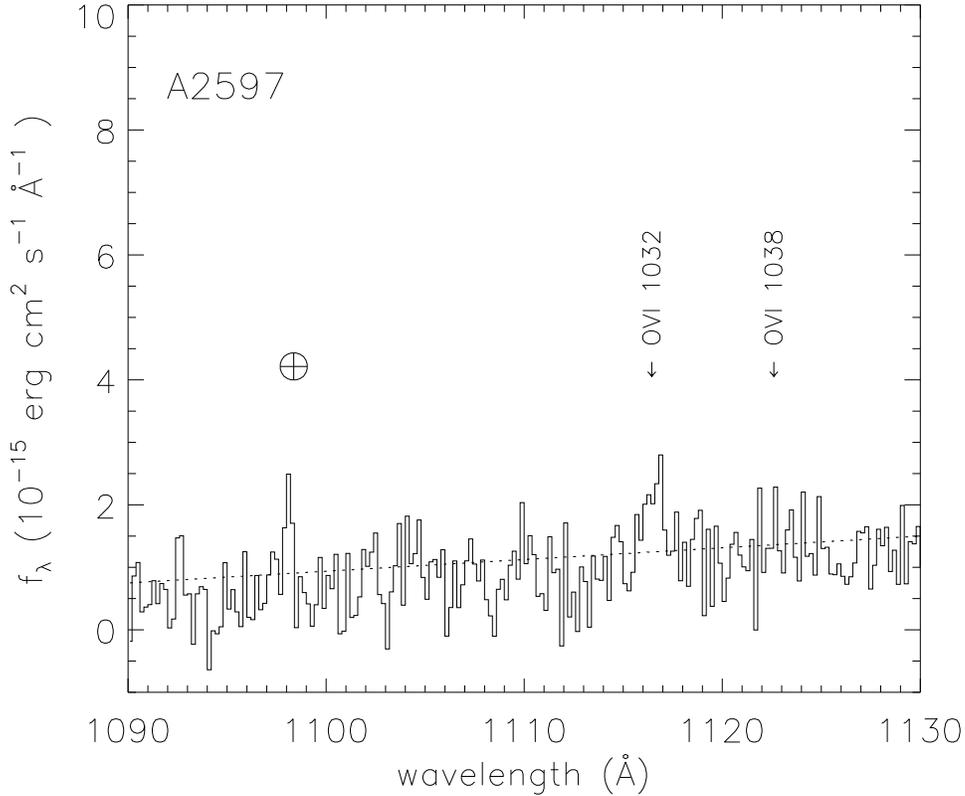}
\vskip 0pt \caption{
{\em FUSE} detection of O~VI in the central
36~kpc of the cooling flow cluster Abell 2597.
The line flux is consistent with the luminosity
expected from $\sim 40 \, M_\odot$ of gas cooling
through $\sim 10^6$~K.  (Figure from Oegerle
et al. 2001.)
\label{fig-fuse_a2597}}
\end{figure*}

High-resolution spectroscopic observations with 
{\em XMM-Newton} and {\em Chandra} are now revealing 
a deficit of emission from gas below $\sim T_{\rm vir} / 3$,
relative to this predicted spectrum.
Peterson et al. (2003) compiled Reflection Grating Spectra (RGS) spectra of 12 
cooling flow clusters, the single largest collection 
to date.  We plot an example from the Perseus cluster in
Figure~\ref{fig-xmm-perseus}. None of the clusters hotter than 4~keV show 
evidence for Fe XVII emission from gas below 1~keV, and
Fe XVII is weaker than expected in cluster with global
temperatures of 2--4~keV.  This line does appear in the 
spectra of supernova remnants, so its absence in cluster 
spectra is not a shortcoming of the plasma codes or the detectors. 
Furthermore, the early {\em XMM-Newton} RGS results (Peterson et al. 2001) 
have been confirmed 
by {\em Chandra} grating spectroscopy (e.g., Hicks et al. 2002).
Gas at $\lesssim 1$~keV apparently does not exist in the
amounts predicted by simple cooling flow models.
Even the data from instruments with lower spectral resolution,
such as the ACIS-S detector on board {\em Chandra},
suggest significantly lower mass cooling rates than obtained from previous 
analyses of {\em ROSAT} and {\em ASCA} data (e.g., McNamara et al. 2000;
Wise \& McNamara 2001; Lewis, Stocke, \& Buote 2002).
Faint detections and strong limits on O~VI emission from the {\em FUSE}
satellite (Oegerle et al. 2001) also imply lower mass cooling
rates (Fig.~\ref{fig-fuse_a2597}).

Two {\it ad hoc}\ models for cool gas do fit the high-resolution 
observations obtained with the {\em XMM-Newton} RGS instrument reasonably well
(Kaastra et al. 2001; Peterson et al. 2001, 2003).  
One is a two-temperature model, in which some gas is at $T_{\rm vir}$ 
and some is at $\sim T_{\rm vir}/2$. The other is a modified 
cooling flow model, in which the amount of cooling gas tapers off from 
$T_{\rm vir}$ to a minimum temperature 
$\sim T_{\rm vir}/3$ (Peterson et al. 2003). 
Because the temperature floor in these models seems to scale 
with $T_{\rm vir}$, it would appear that whatever
prevents the gas from cooling further is sensitive 
to the depth of the cluster potential.

The assumption that cooling flows contain inhomogeneous,
multiphase gas, as implied by their surface brightness profiles,
has also been called into question.  {\em XMM-Newton} observations of M87, 
at the center of the nearest cooling flow cluster, indicate 
that the surrounding intracluster medium consists of a
single temperature plasma, except for those regions 
of the cluster associated with the M87 radio source
(Matsushita et al. 2002).

\subsection{Time for a New Name}

What should we call these clusters in which gas no longer
appears to be cooling and flowing?  The close association
between short central cooling times, H$\alpha$ nebulosity,
and H$_2$ emission strongly suggests that something unusual
is happening in their cores.  Star formation in some cases
is rapid enough to qualify as a starburst (e.g., 
McNamara \& O'Connell 1992; Cardiel et al. 1995), even though
it cannot solve the mass-sink problem.  The goings-on in the
cores of these clusters certainly qualify
as an important astrophysical puzzle that may have far reaching
implications for galaxy formation.  However, as we search for
a new name for ``cooling flow'' clusters, we should perhaps
settle for an observable, such as ``cool core'' clusters,
as has been also suggested by others (Molendi \& Pizzolato 2001).

Adopting a name less freighted with theoretical assumptions
might promote more balanced consideration of alternatives
to the cooling flow hypothesis.  Any successful
model must explain the following features of cool core
clusters:
\begin{itemize}
 \item The apparent lack of a mass sink comparable to
       $\dot{M}_X H_0^{-1}$.
 \item The positive core temperature gradients extending
       to $\sim 10^2$~kpc in clusters with $t_c < H_0^{-1}$.
 \item The frequent incidence of emission-line nebulae,
       dust lanes, and molecular gas in clusters with
       $t_c < H_0^{-1}$ and their absence in clusters
       with $t_c > H_0^{-1}$.
 \item The tendency for radio sources to be present in
       clusters with $t_c < H_0^{-1}$.
\end{itemize}
In light of the new X-ray observations, many of the competing 
ideas that have previously received less attention and testing 
than the cooling flow hypothesis are now being revisited.
The next section discusses how feedback from supernovae
and AGNs might limit the amount of gas that condenses
in clusters, and the following section outlines the
potentially important role of electron thermal
conduction.

\section{The Galaxy-Cluster Connection}

Simple cooling flows may be disproven, but cooling in
general plays a major role in determining the global X-ray
properties of clusters.  Cosmological models of cluster
formation that do not include radiative cooling and the
ensuing feedback processes fail to produce realistic
clusters (Lewis \etal 2000; Pearce et al. 2000; Muanwong
\etal 2001; Voit \& Bryan 2001).  The most glaring failure is in
predictions of the $L_X$-$T_X$ relation.  Models
without galaxy formation predict $L_X \propto T_X^2$
(Kaiser 1986; Borgani \etal 2001; Muanwong \etal 2001), 
while observations indicate $L_X$ \simpropto\ 
$T_X^{3}$ (Mushotzky 1984; Edge \& Stewart 1991; 
David et al. 1993; Markevitch 1998; 
Arnaud \& Evrard 1999; Novicki, Sornig, \& Henry 2002).  
Ignoring cooling and feedback also causes problems
with the slope and normalization of the $M_{\rm vir}$-$T_X$
relation between virial mass and temperature (Horner,
Mushotzky, \& Scharf 1999; Nevalainen, Markevitch, \& Forman 2000;
Finoguenov, Reiprich, \& B\"ohringer 2001), which is
a fundamental ingredient in efforts to constrain cosmological
parameters with cluster observations. 

Recent work has shown that tracing the development
of intracluster entropy is a powerful way to understand
how cooling, supernova feedback, and perhaps energy
injection by AGNs conspire to determine both the $L_X$-$T_X$
and $M_{\rm vir}$-$T_X$ relations of present-day clusters 
(Ponman, Cannon, \& Navarro  1999; Bryan 2000; Voit \& Bryan 2001; 
Voit et al. 2002, 2003; Wu \& Xue 2002a,b).  
Here we briefly outline some
connections between a cluster's galaxies and
its intracluster medium and show how these
connections manifest themselves in the intracluster 
entropy distribution. Then we focus on some particular
models for how AGNs might quench cooling in clusters.

\subsection{The Theoretical Cooling Flow Problem}

\begin{figure*}[t]
\includegraphics[width=1.00\columnwidth,angle=0,clip]{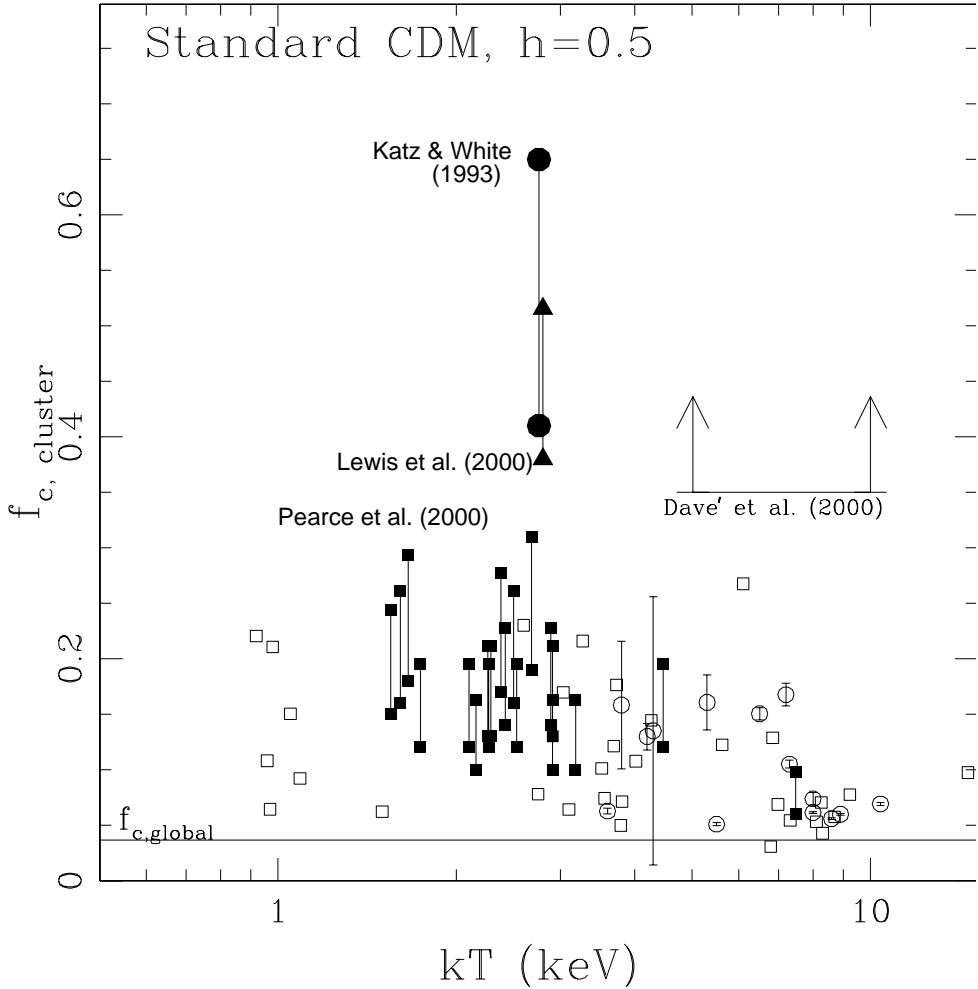}
\vskip 0pt \caption{
          The global overcooling problem.  High-resolution
          cosmological simulations including cooling, represented
          by the labeled solid points, predict that at least 20\%
          of the Universe's baryons should have condensed into stars or 
          cold clouds, if feedback is ineffective.  However, the global
          condensed baryon fraction $f_{\rm c,global}$ inferred
          from large-scale surveys is $\sim$5\%--10\%, depending on
          the initial mass function, and the condensed baryon
          fractions inferred from cluster observations (empty
          squares) are $\sim$10\%--20\%.  (Figure from Balogh
          et al. 2001.)
\label{fig-condfrac}}
\end{figure*}

Cosmological models for cluster formation that do not
include cooling are clearly too simplistic because they
do not spawn galaxies.  Radiative cooling initiates
galaxy birth but is responsible for the now-classic
overcooling problem (White \& Rees 1978; Cole 1991).
If no form of feedback opposes cooling, then at least
20\% of the baryons in the Universe, and maybe more,
should have condensed into stars.  Yet, the observed
fraction of baryons in stars is $\lesssim 10$\% 
(see Fig.~\ref{fig-condfrac}; Balogh et al. 2001).
This overcooling problem is even more acute in clusters,
where primordial densities are higher, enabling even
more of the baryons to condense.

One could also call this problem the ``theoretical
cooling flow problem'' because far too many baryons
cool and condense if there is no heat source to compensate
for radiative cooling.  Supernova feedback is generally
assumed to provide the requisite heat to halt overcooling in galaxies, 
although the precise mechanism remains murky (e.g., Kay \etal 2002).
However, supernovae might not provide enough heat
to halt overcooling in clusters, where the binding
energy per particle exceeds the mean supernova energy
per particle ($\sim 1$~keV), as measured from the
intracluster metallicity (e.g., Finoguenov, Arnaud, 
\& David 2001).  Thus, feedback from AGNs may be necessary 
to suppress cluster cooling flows.

\subsection{Cooling, Feedback, and Intracluster Entropy}

The slope of the observed $L_X$-$T_X$ relation has long been
assumed to reflect an early episode of feedback that
imposed a universal entropy floor throughout the
intergalactic medium (Evrard \& Henry 1991; Kaiser 1991).
An entropy floor steepens the $L_X$-$T_X$ relation
from $L_x \propto T_X^2$ to $L_X$ \simpropto\ $T_X^{3}$
because the extra entropy stiffens the intracluster
medium against compression.  Lower temperature clusters
with shallower potential wells therefore have a harder
time compressing their core gas, leading to lower
core densities and smaller X-ray luminosities than
expected in models without cooling and feedback.

Measurements of intracluster entropy in the vicinity 
of the X-ray core radius support this notion because they 
indicate elevated entropy levels in groups and poor clusters, 
corresponding to $Tn_e^{-2/3} \approx 100 - 150 \, 
{\rm keV \, cm^2}$ (Ponman et al. 1999;  Lloyd-Davies, Ponman, \& Cannon
2000).  In order to produce such an
entropy floor through supernova heating alone,
a large proportion of the available supernova
energy is needed (Kravtsov \& Yepes 2000).  
Even then, the required supernova heating 
efficiency may be unrealistic, 
in which case additional heat input from
AGNs would be required (Valageas \& Silk 1999; 
Wu, Fabian, \& Nulsen 2000).  However, there is another 
way to interpret these entropy measurements that 
does not involve global heating of the 
intergalactic medium.

\begin{figure*}[t]
\includegraphics[width=1.00\columnwidth,angle=0,clip]{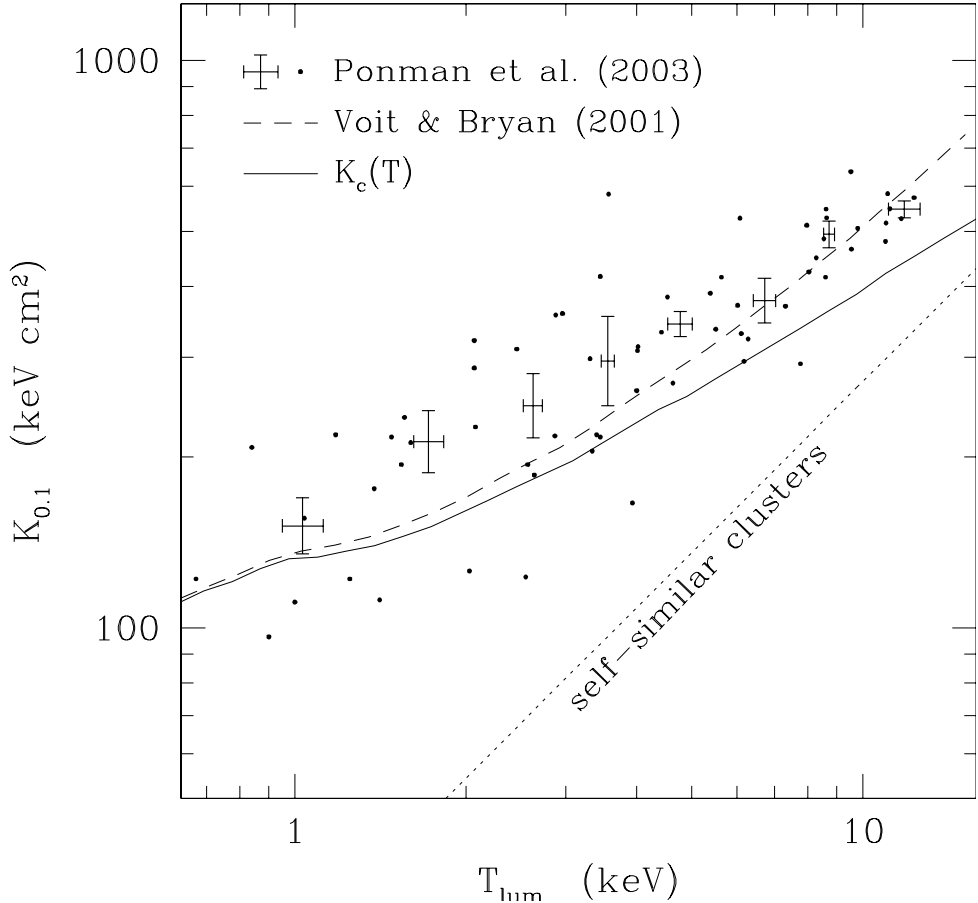}
\vskip 0pt \caption{
Relationship between core entropy and the cooling threshold.  
Each point with error bars shows the mean core entropy $K_{0.1}$, 
measured at $0.1 r_{200}$, for eight clusters within a given 
bin of luminosity-weighted temperature $T_{\rm lum}$, and small 
circles show measurements for individual clusters (Ponman \etal 2003).
The dotted line shows a self-similar relation calibrated using 
the median value of $K_{0.1}$ measured in simulation L50+ of 
Bryan \& Voit (2001), which does not include cooling or feedback.
The solid line shows the cooling threshold $K_c(T)$, defined 
to be the entropy at which the cooling time equals 14~Gyr,
assuming the cooling function of Sutherland \& Dopita (1993) 
for 0.3 solar metallicity.  The dashed line shows the entropy
at $0.1r_{200}$ in the model of Voit \& Bryan (2001) when this cooling
function is used.  
\label{ent_core}}
\end{figure*}

Instead, the $L_X$-$T_X$ relation may reflect a
conspiracy between cooling and feedback that regulates
the core entropy of clusters and groups (Voit \& Bryan 2001).
Figure~\ref{ent_core} shows measurements of core
entropy from Ponman, Sanderson, \& Finoguenov  (2003)
along with the locus in $T$-$Tn_e^{-2/3}$ space
at which the cooling time equals a Hubble time.
The way in which core entropy tracks this locus
suggests that gas with a short cooling time
is eliminated from clusters by a combination
of cooling and feedback.

A parcel of gas with entropy ($Tn_e^{-2/3}$) below this 
threshold must condense unless feedback intervenes.  
If feedback is effective, then it will raise the entropy 
of the gas parcel until it exceeds the threshold, 
where it is no longer subject to cooling.  If feedback 
is ineffective, then most of the parcel's gas will 
cool and condense.  Either way, both cooling and 
feedback remove gas from the region below the threshold, 
establishing a core entropy at the level of the threshold.

This mechanism explains why simulations that include
cooling produce clusters with reasonably realistic
$L_X$-$T_X$ and $M_{\rm vir}$-$T_X$ relations, regardless of the efficiency 
of feedback (Muanwong \etal 2001; Borgani \etal 2002;
Dav\'e, Katz, \& Weinberg 2002).
However, the amount of baryons that end up in galaxies
is very sensitive to how feedback is implemented 
(Kay, Thomas, \& Theuns 2003).
Thus, it would appear that cooling is essential
to a proper understanding of cluster properties
and that the details of how cooling flows are
suppressed are crucial to understanding hierarchical
galaxy formation in the context of clusters.

\subsection{AGNs and Cooling Flows}

Many cooling flow clusters also contain radio sources indicative
of recent nuclear activity (e.g., Burns 1990).  This close association 
between AGNs and clusters with short central cooling times
supports the idea that feedback from AGNs helps to suppress
cooling.  Some authors have proposed that radiation
from the active nucleus heats the cluster core (e.g., Ciotti
\& Ostriker 1997, 2001), but far more attention has been paid
to the possibility that radio jets somehow heat the
intracluster medium (e.g., Binney \& Tabor 1995; Churazov \etal 2001;
Soker \etal 2001; Br\"uggen \& Kaiser 2002; Reynolds, Heinz, \& Begelman 2002). 
Such heating was originally not considered to be a 
viable solution to the mass-sink problem because 
the total amount of energy needed to stabilize 
a strong cooling flow is quite large ($\sim 10^{62}$~erg),
and the spatial deposition of that heat would need to be 
precisely matched to local cooling rates in order to
maintain thermal stability (Fabian 1994).  However,
{\em Chandra} and {\em XMM-Newton} observations showing
widespread interactions between radio plasma and the
intracluster medium (e.g., Fabian \etal 2000; McNamara \etal 2000)
have stimulated new interest in connections between
radio jets and cooling flows.

The high spatial resolution of the {\em Chandra} observations 
reveals that jets do not simply shock-heat the surrounding 
intracluster medium, because the gas surrounding the lobes 
appears somewhat cooler and denser than the undisturbed
gas farther from the lobes (McNamara et al. 2000).  
Thus, because cluster cores do not 
appear to be shock heated, most of the recent theoretical 
models have focused on mixing and turbulent heating stirred 
up as the buoyant radio plasma rises through the intracluster 
medium (e.g., Quilis, Bower, \& Balogh 2001; Br\"uggen \& Kaiser 2002; 
Reynolds et al. 2002).  Both mixing and heating raise the entropy of the core
gas, consequently raising its cooling time as well.
These models circumvent the local fine-tuning problem 
by distributing heat over a large region 
through convection, and they add additional
thermal energy to the core beyond that supplied by the
AGN itself by mixing the core gas with
overlying gas of higher entropy.

However, not all clusters with short central cooling
times have obvious nuclear activity.  Thus, if
AGN heating is the solution to the cooling flow
puzzle, then it must be episodic.  A recent model
by Kaiser \& Binney (2003) shows how the central
entropy profile would evolve under episodic
heating.  Because cooling rates rise dramatically
as isobaric gas cools to lower temperatures,
an episodically heated medium usually contains
very little gas below $\sim T_{\rm vir}/3$.
When the central gas reaches this temperature
it is assumed to cool very quickly to even cooler 
temperatures and accrete onto the AGN, triggering
another episode of heating.  This feature of
episodic heating may explain the absence of
line emission from colder gas in cool core
clusters.

\section{The Revival of Conduction}

During the first two decades of the cooling flow hypothesis,
the idea that electron thermal conduction might somehow suppress
cooling was a minority viewpoint, despite the fact that it
has many attractive features.  Because conduction carries
heat from warmer regions to cooler regions, it naturally
directs thermal energy into regions that would otherwise
condense.  Also, it taps the vast reservoir of thermal
energy in the intracluster medium surrounding the cluster
core, which is more than sufficient to resupply the radiated
energy.  

Many models invoking conduction have been developed
(e.g., Tucker \& Rosner 1983; Bertschinger \& Meiksin 1986;
Bregman \& David 1988; Rosner \& Tucker 1989; Sparks 1992), but 
conduction has often been dismissed as a global solution
on the grounds that it is not stable enough to preserve 
the observed temperature and density gradients for periods
of order $\gtrsim 1$~Gyr (Cowie \& Binney 1977; Fabian 1994).
The heat flux from unsaturated conduction proceeding
uninhibited by magnetic fields is $\kappa_s \nabla T$,
with $\kappa_s \approx 6 \times 10^{-7}\,T^{5/2}\, {\rm erg \, cm^{-1} \, 
s^{-1} \, K^{-7/2}}$, the so-called Spitzer
rate (Spitzer 1962).  Because of this extreme sensitivity
to temperature, it is difficult for radiative cooling and
conduction to achieve precise thermal balance with a globally
stable temperature gradient (Bregman \& David 1988; Soker 2003).  
However, any mechanism that places cool gas at the
center of a cluster, such as a merger of a gas-rich galaxy 
with the central cluster galaxy, sets up a temperature 
gradient that would cause uninhibited conduction 
to proceed until either the cool gas has evaporated or 
the hot gas has condensed (Sparks \etal 1989).
As long as a temperature gradient exists, a certain
amount of conduction has to occur.

In order for a standard, steady cooling flow alone to produce the
temperature gradients observed in cool core clusters, conduction
must be highly suppressed by at least 2 orders of magnitude
below the Spitzer rate, presumably by tangled magnetic fields
(Binney \& Cowie 1981; Fabian \etal 1991).  
Yet, recent theoretical analyses of conduction have concluded
that this level of suppression is unrealistically high
(Malyshkin 2001; Malyshkin \& Kulsrud 2001; Narayan \&
Medvedev 2001).  These studies suggest that magnetic field
tangling may only suppress conduction by a factor $\sim$3--10,
implying that it may be important in the cores of clusters.

\begin{figure*}[t]
\hspace{-0.5cm}
\includegraphics[width=0.90\columnwidth,angle=-90,clip]{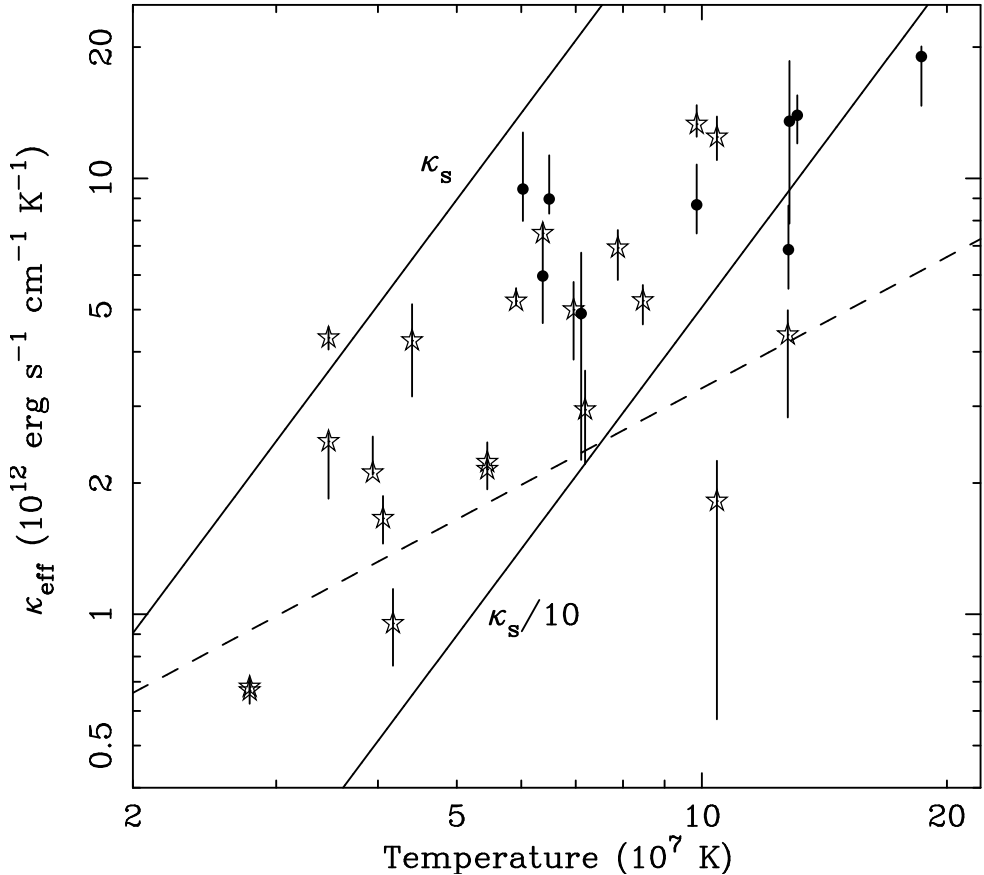}
\vskip 0pt \caption{
          Effective conduction coefficients $\kappa_{\rm eff}$
          required for conduction to compensate for radiative
          cooling within the central regions of clusters,
          plotted as a function of cluster temperature.
          The required conductivity generally does not
          exceed the Spitzer rate $\kappa_{\rm S}$
          at radii $\sim 100$~kpc, implying that
          conduction is potentially important in
          cluster cores.  (Figure from Fabian \etal 2002.)
\label{fig-kappaeff}}
\end{figure*}

This finding, coupled with the X-ray spectroscopic observations 
showing little evidence for cooling gas, has helped to spur
a remarkable revival of the idea of conduction, 
with notable assistance from some of its
harshest earlier critics (Fabian et al. 2002;
Voigt \etal 2002; Zakamska \& Narayan 2003; but see Loeb 2002). 
One can analyze the observed temperature gradients of
clusters by defining an effective conduction coefficient
$\kappa_{\rm eff}(r) \equiv L(<r)/4 \pi r^2 (dT/dr)$ 
that would lead to balance between radiative cooling and 
conductive heating.  The values of $\kappa_{\rm eff}$
measured at $\sim 100$~kpc in cool core clusters are
typically $\sim (0.1-0.3) \kappa_{\rm S}$, suggesting
that electron thermal conduction is a plausible mechanism
for counteracting radiative cooling over much of the region
where $t_c < H_0^{-1}$ (Fig.~\ref{fig-kappaeff}).

Even though conduction may be important at $\sim 100$~kpc,
the required effective conductivity exceeds the Spitzer
rate at radii $\sim 10$~kpc (Ruszkowski \& Begelman 2002; Voigt \etal 2002), 
a result presaged by the analysis of 
Bertschinger \& Meiksin (1986).  Thus, a modest amount
of feedback may be necessary to offset cooling in the
centers of cool core clusters.  Hybrid models involving
conduction in the outer parts of the core and AGN feedback
in the inner parts have been developed by Ruszkowski
\& Begelman (2002) and Brighenti \& Mathews (2003).

\section{Paths to a Resolution}

Observations from the present generation of X-ray telescopes
have dethroned the cooling flow hypothesis, but what will
take its place?  Star formation, radio jets, and conduction
may all have important roles to play in the development
of cluster cores.  Conduction is notoriously hard to test
because the rate at which it proceeds depends on the unknown
geometry of intracluster magnetic fields and uncertain factor
by which these fields suppress heat flow.  Looking for hallmarks
of episodic feedback, from both AGNs and supernovae, may be
more fruitful, at least in the short term.

If feedback is episodic, then the state of the central
intracluster medium should be closely related to other
goings-on in the cluster core.
Thus, it would be interesting to test whether
the $\sim T_{\rm vir}/3$ scaling of the minimum plasma temperature
apparent in the early sample of {\em XMM-Newton} clusters from 
Peterson et al. (2003) holds for a large sample
of cool core clusters with various levels of core activity.
How do the X-ray emission-line spectra of clusters with 
radio-loud nuclei differ from those of clusters with radio-quiet
nuclei?  Are there any correlations between X-ray line emission
and the presence of obvious star formation or emission-line nebulae? 
Episodic heating also leads to a predictable pattern in
the evolution of the core entropy distribution (Kaiser \&
Binney 2003).  Thus, studying the core entropy distributions
of a large sample of clusters may reveal a telltale pattern
of entropy evolution with time.  

In order to look for evidence of a feedback duty cycle
in cluster cores and to study how their properties depend
on AGN and star formation activity, we are now in the midst
of an archival {\em Chandra} study of cluster cores.
The result of this program will be a publicly available 
library of entropy distributions showing how the entropy
of intracluster gas depends on radius and enclosed gas
mass within that radius (Horner et al., 
in preparation.)  We are focusing on entropy because 
it is the thermodynamic quantity most closely related 
to heat input and radiative cooling.  We invite all who 
are interested in the vexing problem of cooling flows
to take advantage of this database.

\end{document}